\documentclass[draft]{agujournal2019}
\usepackage{url} 
\usepackage{lineno}
\usepackage[finalnew]{trackchanges}
\usepackage{soul}

\usepackage{verbatim}

\linenumbers
\draftfalse

\journalname{Geophysical Research Letters}

\soulregister\ref7
\soulregister\cite7
\begin{document}
\nolinenumbers

%
%


\title{Magnetic Field and Plasma Asymmetries Between the Martian
Quasi-Perpendicular and Quasi-Parallel Magnetosheaths}

%
%




\authors{Abigail Tadlock\affil{1}, Chuanfei Dong\affil{1}, Chi Zhang\affil{1}, Markus Fränz\affil{2}, Hongyang Zhou\affil{1}, Jiawei Gao\affil{1}}

 \affiliation{1}{Center for Space Physics and Department of Astronomy, Boston University, Boston, MA, USA}
  \affiliation{2}{Max‐Planck‐Institute
for Solar System Research, Göttingen, Germany}




\correspondingauthor{Chuanfei Dong}{dcfy@bu.edu}
\correspondingauthor{Chi Zhang}{zc199508@bu.edu}


\begin{keypoints}
\item Magnetic fields and plasmas exhibit clear asymmetrical distributions in the Martian magnetosheath.
\item The asymmetries in the Martian magnetosheath exhibit similarities and differences compared to those in Earth's and Venus’ magnetosheaths.
\item The Martian magnetosheath is influenced by several factors, including current sheet structures, planetary ions, and shock geometry.
\end{keypoints}
%
%

%
%


\begin{abstract}
The Martian magnetosheath acts as a conduit for mass and energy transfer between the upstream solar wind and its induced magnetosphere. However, our understanding of its global properties remains limited. Using nine years of data from NASA's Mars Atmosphere and Volatile EvolutioN (MAVEN) mission, we performed a quantitative statistical analysis to explore the spatial distribution of the magnetic fields, solar wind and planetary ions in the magnetosheath. We discovered significant asymmetries in the magnetic field, solar wind protons, and planetary ions between the quasi-perpendicular and quasi-parallel magnetosheaths. The asymmetries in the Martian magnetosheath exhibit both similarities and differences compared to those in the Earth's and Venus' magnetosheaths. These results indicate that the Martian magnetosheath is distinctly shaped by both shock geometry and planetary ions.

\end{abstract}

\section*{Plain Language Summary}
The magnetosheath, located between the bow shock and the planetary (induced) magnetosphere, plays a crucial role in transferring mass and energy from the upstream solar wind to the upper atmosphere. Knowing its characteristics is essential to understand the interactions of the solar wind with planetary systems. Here, we conducted a statistical analysis to explore the global properties of magnetic fields, solar wind protons, and planetary ions within the Martian magnetosheath. Our findings reveal a distinct asymmetric distribution of the magnetic field, solar wind and planetary ions between the quasi-perpendicular and quasi-parallel magnetosheaths. Our results indicate that the properties of the Martian magnetosheath are influenced by a combination of factors. The Martian magnetosheath provides a unique and natural laboratory to study plasma physics processes involving both magnetized solar wind and planetary heavy ions.

%
%

 \section{Introduction}
%


The interaction of the solar wind with Mars is widely recognized as a crucial factor in the long-term evolution of the Martian atmosphere \cite<e.g.,>{luhmann_solar_1992,lillis_characterizing_2015,Dong2018ApJL,jakosky_loss_2018, Zhang2025a}. Before this interaction occurs, the properties of the pristine solar wind are modified by the bow shock (BS). Specifically, as the super-Alfv\'{e}nic solar wind, carrying the interplanetary magnetic field (IMF), approaches the planetary magnetosphere, it crosses a bow shock that decelerates and heats it \cite{nagy_plasma_2004,mazelle_bow_2004,Zhang2025b}. This decelerated solar wind forms a layer known as the magnetosheath, which envelops the magnetosphere \cite{bertucci_induced_2011,narita_magnetosheath_2021}. Thus, the magnetosheath plays a pivotal role as a conduit for the transfer of mass and energy between the upstream solar wind and the planetary magnetosphere, whether intrinsic or induced \cite<see>[for review]{lucek_magnetosheath_2005}.

Within our solar system, the Martian magnetosheath serves as a unique laboratory among terrestrial planets for studying the interactions between the solar wind and its induced magnetosphere. Note that some comets exhibit similar characteristics \cite{He2021,goetz_plasma_2022}. This uniqueness is primarily due to three distinct factors. First, the magnetic field environment of Mars is hybrid, characterized by both induced magnetic fields and localized crustal magnetic fields \cite{nagy_plasma_2004,Luhmann2015GRL,dong_solar_2015,dibraccio_twisted_2018,zhang_three-dimensional_2022,Zhang2023,Gao2024}. This combination results in a considerably more complex magnetic environment compared to that of other planets. Second, the Martian magnetosheath contains more planetary ions compared to those of other planets \cite{Xu2016}, which significantly influence its dynamics, including localized turbulence and wave activity  \cite{li_pickup_2024}. This elevated ion population is largely attributed to Mars' extended exosphere, which expands and contracts seasonally due to the planet’s low gravity and highly elliptical orbit \cite{valeille_three-dimensional_2009,dong_solar_2018}. Third, the relatively small thickness of the magnetosheath, comparable to the gyroradius of solar wind protons, suggests that the solar wind is not fully thermalized \cite{moses_expectations_1988}, emphasizing the importance of kinetic effects at Mars. These unique attributes make the Martian magnetosheath a particularly interesting area of study.

To date, research on the Martian magnetosheath has explored several key aspects, including waves and turbulence \cite{russell_upstream_1990,sagdeev_wave_1990,dubinin_martian_1997,espley_low-frequency_2005,bertucci_induced_2011, ruhunusiri_low-frequency_2015, Dong2015TESS,dubinin_ultra-low-frequency_2016, li_pickup_2024, romanelli_alfven_2024,simon_wedlund_statistical_2023, simon_wedlund_local_2025,Zhang2025c}, electron energization \cite{horaites_observations_2021}, temperature anisotropy \cite{halekas_flows_2017, andreone_properties_2022}, and the deflection of solar wind protons \cite{dubinin_solar_2018, romanelli_variability_2020}. However, compared to our understanding of Earth's magnetosheath, our knowledge of the Martian magnetosheath remains limited. Previous studies have clearly shown that the properties of the Earth's magnetosheath are strongly dependent on the shock geometry \cite{dimmock_dawn-dusk_2016, dimmock_dawn-dusk_2017}. 

The bow shock can be classified into two types based on geometry: the quasi-parallel shock ($Q_\parallel$) and the quasi-perpendicular shock ($Q_\perp$). A $Q_\parallel$ ($Q_\perp$) shock occurs where the IMF is approximately parallel (perpendicular) to the bow shock normal \cite{bale_quasi-perpendicular_2005, burgess_quasi-parallel_2005}. The downstream magnetosheath in the $Q_\parallel$ shock, called the $Q_\parallel$ sheath, exhibits properties distinct from those of the $Q_\perp$ sheath. For example, \citeA{dimmock_dawn-dusk_2016, dimmock_dawn-dusk_2017} and \citeA{walsh_dawn-dusk_2012} showed that the proton density and ion temperature were higher on the $Q_\parallel$ flank at Earth, whereas magnetic field strength $|B|$, proton speed, and ion temperature anisotropy were larger on the $Q_\perp$ flank. However, the asymmetries in the Venusian magnetosheath exhibit similarities and differences compared to those observed on Earth, suggesting that induced magnetosheaths may differ from intrinsic magnetosheaths \cite{rojas_mata_proton_2023}. At Mars, several asymmetric properties have also been noted, including the sheath thickness \cite{halekas_flows_2017}, electron temperature \cite{andreone_properties_2022}, and the magnetic field draping pattern \cite{zhang_three-dimensional_2022}. Despite this, the overall properties of asymmetry in the Martian magnetosheath remain unclear. Several key questions still remain: Does the Martian magnetosheath exhibit significant asymmetries in the distribution of magnetic fields and plasma, including both solar wind protons and planetary ions? If so, how pronounced are these asymmetries, and how do they compare to those observed at other planets? Additionally, what factors control these asymmetries? Addressing these questions can significantly enhance our understanding of the physics occurring within planetary magnetosheaths and the interactions between the solar wind and planets.

To address the questions raised, we conducted a statistical analysis of plasma and magnetic field distributions in the Martian magnetosheath, utilizing nine years of MAVEN observations. Our goal was to quantitatively reveal the asymmetries in the magnetic field and plasma present in this region.

 \section{Data}
%
%

 To perform this study, we used magnetic field data from the magnetometer (MAG) \cite{connerney_maven_2015} heavy and ion data from the Suprathermal and Thermal Ion Composition (STATIC) instrument \cite{mcfadden_maven_2015}, and proton data from the Solar Wind Ion Analyzer (SWIA) \cite{halekas_solar_2015} onboard MAVEN. We used Mars-Solar-Electric (MSE) coordinates. In MSE coordinates, the $x$-axis points toward the Sun, the $z$-axis aligns with the solar wind electric field, $\vec{E}_{sw} = -\vec{v}_{sw} \times \vec{B}_{IMF}$, and the $y$-axis completes the right-handed coordinate system. Following \citeA{zhang_three-dimensional_2022} and \citeA{zhang_energetic_2024}, we only construct the MSE coordinates for orbits with steady IMF conditions, characterized such that the angle between the inbound and outbound IMF is less than 30°, as measured in-situ with MAG. Between December 2014 and December 2023, we found 6338 orbits with available plasma data, magnetic field data, and associated MSE coordinates.
 
We used plasma moments of $O^+$ and $O_2^+$ ions using data from STATIC measurements, applying the integral method described in \citeA{franz_plasma_2006}, \citeA{zhang_mars-ward_2022}, \citeA{zhang_energetic_2024}, and \citeA{fowler_-situ_2022}.

 \section{Method}
\begin{figure}
    \centering
    \includegraphics[width=1\linewidth]{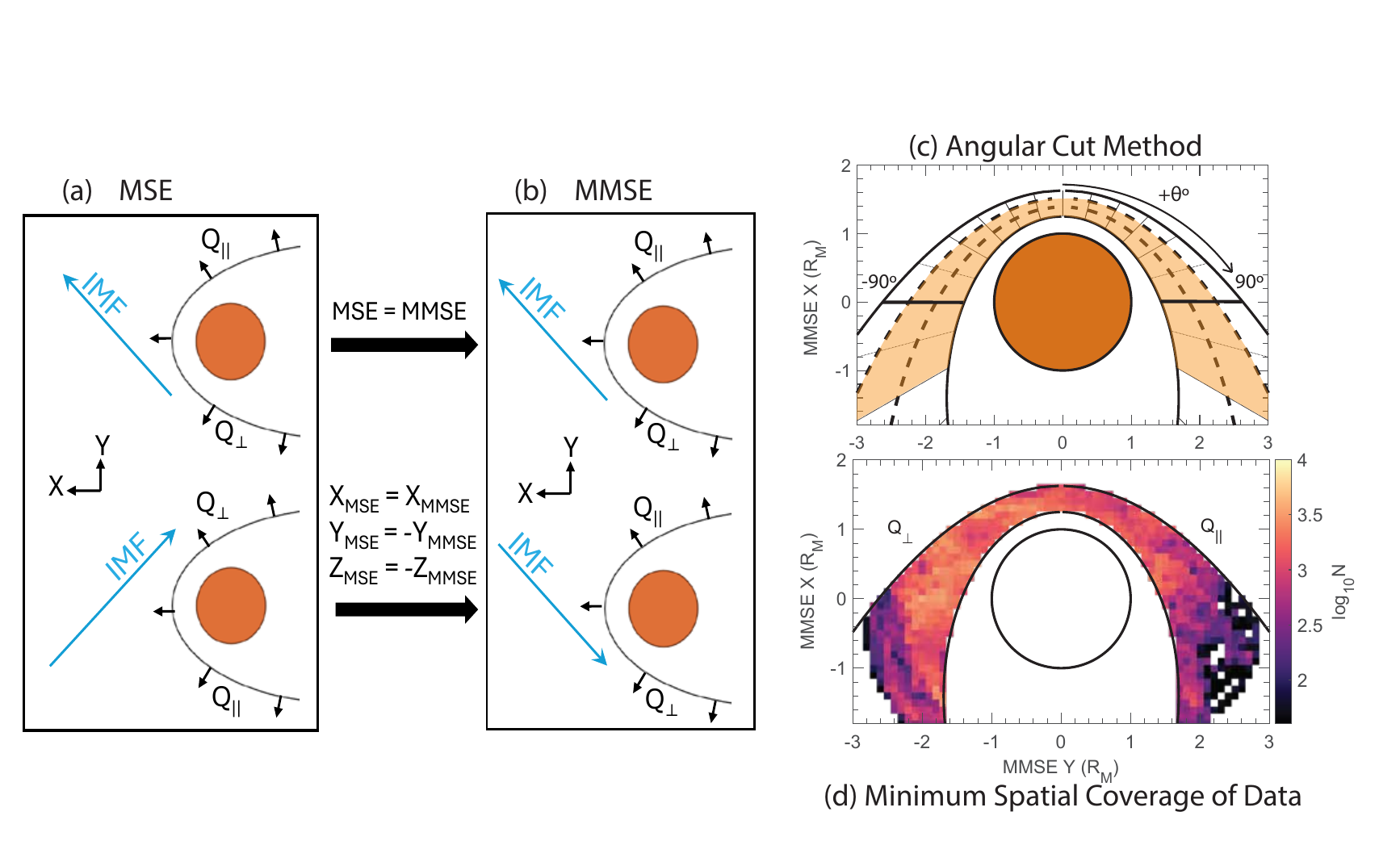}
    \caption{a) Two example IMF configurations in MSE coordinates. The IMF has a +$B_{x}$ component in the top half panel and a -$B_{x}$ component in the bottom half panel. The BS (black curve) is drawn with various vectors indicating the shock normal at different locations. The $Q_\parallel$ and $Q_\perp$ shocks are labeled. b) The conversion of both configurations to MMSE coordinates. c) A diagram of our angular cut method. The measurement area is shown as the orange shaded region, with $15^\circ$ bin intervals (the bin overlap is not shown). d) The spatial coverage of our data. $N$ is the bin count. This map is for STATIC ($O^+$, $O_2^+$), which has the most limited data coverage.}
    \label{fig:mmse}
\end{figure}

In the MSE coordinate system, the $Q_\parallel$ magnetosheath is located in the +$Y$ hemisphere when the IMF $B_{x}$ is positive and in the -$Y$ hemisphere when $B_{x}$ is negative. In contrast, the $Q_\perp$ magnetosheath is positioned in the opposite hemisphere (see Figure \ref{fig:mmse}(a)). Thus, to facilitate a unified analysis regardless of the IMF $B_{x}$ polarity, we introduce a new coordinate system called Magnetosheath-MSE (MMSE) coordinates. When IMF $B_{x}$ is positive, MMSE coordinates align with MSE coordinates. For negative IMF $B_{x}$, $Y_{MMSE}$ and $Z_{MMSE}$ are inverted relative to $Y_{MSE}$ and $Z_{MSE}$, ensuring that the $Q_\parallel$ ($Q_\perp$) flank is consistently located in the $+Y_{MMSE}$ ($-Y_{MMSE}$) flank (see Figure \ref{fig:mmse}(b)).

To analyze the asymmetries between the $Q_\parallel$ and $Q_\perp$ magnetosheaths, we organized the data into spatial grid bins sized 0.1 $R_M$ by 0.1 $R_M$ ($R_M$ is the radius of Mars, 3390 km) in the MMSE-$XY$ plane. In order to ensure a smooth transition between features, we overlap the bins by 50\%. Given that the magnetosheath displays asymmetries along the $Z_{MMSE}$ (±$E_{sw}$) direction, particularly the heavy ion escape plume in the +$E_{sw}$ hemisphere, \cite<e.g.>{dong_solar_2014,Dong2015ICME,romanelli_variability_2020, dubinin_solar_2018,dubinin_solar_2024,nesbit-ostman_instantaneous_2025} we limited our analysis to data near the magnetic equator, specifically between -1 and 1 $R_M$ in the $z$-direction, to reduce this influence. The data are averaged along the $z$-axis, yielding one median value in the $XY$-plane. As the location of the $Q_\parallel$ and $Q_\perp$ shocks is caused by the IMF cone angle, defined as the angle between the projected IMF and the $x$-axis, we limited our analysis to data with a corresponding upstream cone angle between $30^\circ-60^\circ$ or $120^\circ-150^\circ$. For statistical reliability, we included only bins that contain more than 40 measurements. Moreover, we excluded data potentially affected by crustal fields by setting $|B_{obs}|>10|B_{model}|$, where $|B_{obs}|$ and $|B_{model}|$ are the observed magnetic field strength and field strength predicted by the latest crustal fields model from \citeA{gao_spherical_2021}, respectively. In addition, since the magnetosheath properties are correlated with the upstream solar wind parameters, we normalized the data from each orbit by its corresponding upstream value. Figure \ref{fig:mmse}(d) shows the spatial coverage and bin counts of our STATIC dataset, which has the most limited data coverage. Spatial coverage maps for MAG and SWIA are provided in the supplemental information.

To quantitatively analyze the asymmetries, we used the BS and magnetic pile-up boundary (MPB) models in the magnetic equator plane (XY plane with $Z_{MMSE}$ = 0) from \citeA{trotignon_martian_2006} to define the magnetosheath region. As the BS and MPB are modeled in three dimensions, they extend further inward on the $XY$ plane in the $+Z_{MMSE}/-Z_{MMSE}$ hemisphere compared to the magnetic equator plane ($Z_{MMSE} = 0$). Consequently, we confined our analysis to the inner two-thirds of the magnetosheath, shown as the shaded orange region in Figure \ref{fig:mmse}(c). This approach ensures that most of the data points are located within the true magnetosheath. This approach also limits the effects of using a shock model rather than individual crossings, by eliminating the region where the \change{shocks}{shock's positions} are averaged over. To divide the magnetosheath into thirds, two inner bounds are defined by linearly interpolating between the MPB and BS models, producing two intermediate boundaries (shown as dashed lines in Figure \ref{fig:mmse}(c)).This suffices for our study of large scale structure, but note that as these models are averaged over many different solar wind conditions, this interpolation is less accurate towards the terminator. Furthermore, in order to understand how asymmetries evolve as the solar wind propagates tailward, we defined the angle $\theta$ as the angle between the position vector and the $+X_{MMSE}$ direction (see Figure \ref{fig:mmse}(c)). We define $\theta$ so that it is positive (negative) in the $Q_\parallel$ ($Q_\perp$) magnetosheath. We then rebinned the data in $15^\circ$ bins ranging from $-120^\circ$ to $120^\circ$, with 50\% overlap to ensure a smooth transition of features. Following \citeA{dimmock_dawn-dusk_2017}, we calculated the median values in each angular bin and then calculated the dimensionless asymmetry parameter, \textit{A}, as follows: \begin{equation}
    A = 100 \times \frac{Q_\parallel - Q_\perp}{Q_\parallel + Q_\perp}
\end{equation}
where $Q_\parallel$ and $Q_\perp$ are the previously calculated averages of the bins spanning the same angle from noon on the $Q_\parallel$ and $Q_\perp$ flanks. We quantified the spread of the data in each bin by calculating the standard error of the mean, defined as $\sigma/\sqrt{N}$, where $\sigma$ is the standard deviation and $N$ the number of data points per bin. This allowed us to determine an upper and lower estimate for \textit{A} \cite<see>{dimmock_dawn-dusk_2017}.  
 
 \section{Results} 

 \begin{figure}
    \centering
    \includegraphics[width=0.99\linewidth]{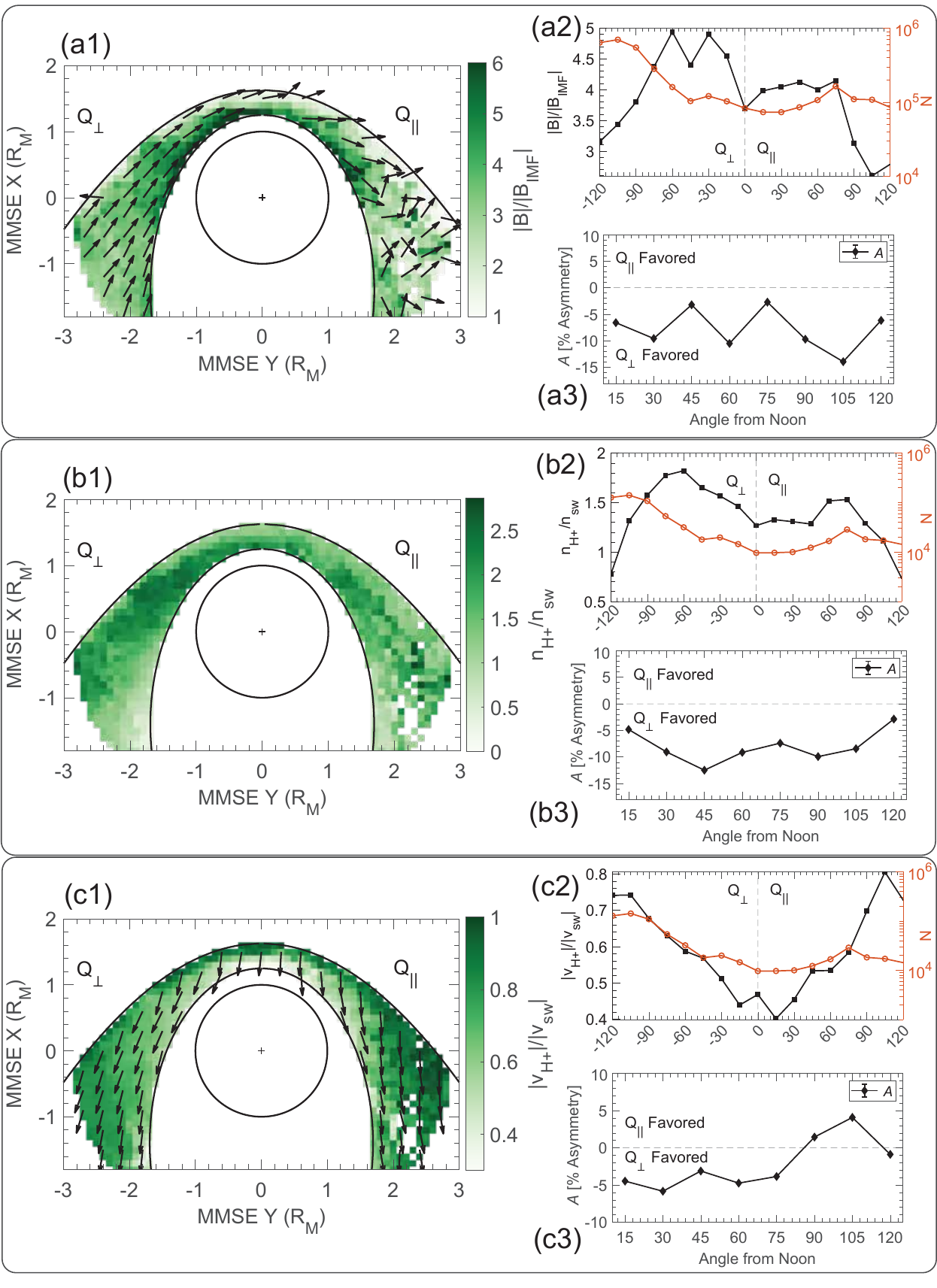}
    \caption{Statistical mapping results for (a1) magnetic field strength with overplotted direction vectors projected into the $XY$ plane, (b1) proton density, and (c1) proton velocity, also with overplotted direction vectors. All values are normalized by upstream conditions. Panels (a2), (b2), and (c2) show the value of the angular bins and the number of data points in each bin $N$. Figures (a3), (b3), and (c3) show the percent asymmetry calculated from these bins.}
    \label{fig:hydrogen}
\end{figure}

\subsection{Magnetic Field Strength}
Figure \ref{fig:hydrogen}(a1) shows the spatial distribution of the magnetic field strength normalized to the upstream IMF strength, denoted $|B|/|B_{IMF}|$. $|B|/|B_{IMF}|$ increases moving from the BS to the MPB. Furthermore, $|B|/|B_{IMF}|$ is higher in the $Q_\perp$ sheath compared to the $Q_\parallel$ sheath, consistent with observations made in Earth's magnetosheath \cite<e.g.,>{dimmock_dawn-dusk_2017}, as well as previous observations at Mars \cite{halekas_flows_2017}. This can be explained by the Rankine–Hugoniot conditions \cite{hudson_discontinuities_1970}: the magnetic field component normal to the BS remains unchanged upon crossing the BS due to the continuity of the normal magnetic field, while the tangential component significantly increases. Therefore, in the $Q_\parallel$ shock, the IMF is dominated by the normal component, resulting in a minimal change in $|B|$ when crossing the shock. In contrast, when the IMF crosses the $Q_\perp$ shock, the $|B|$ is significantly enhanced due to the dominant tangential component.

Figure \ref{fig:hydrogen}(a2) presents the average $|B|/|B_{IMF}|$ in each angular bin. $|B|/|B_{IMF}|$ is larger from noon to $\pm60^o$ and decreases towards past the terminator, caused by the stronger compression near the subsolar region. The asymmetry index, \textit{A}, shown in Figure \ref{fig:hydrogen}(a3), varies from 3\% to 14\%. The area extending beyond $90^\circ$ from local noon in the $Q_\parallel$ region corresponds to the foreshock under the nominal Parker spiral angle of the IMF at Mars (57°) (see Figure 1e of \citeA{jarvinen_ultra-low_2022}), which is characterized by weak and turbulent magnetic fields, as seen in the overlapped direction vectors in Figure \ref{fig:hydrogen}(a1). Consequently, the asymmetry becomes more pronounced. 

In the $Q_\perp$ sheath,  $|B|/|B_{IMF}|$ ranges from 3 to 5, while in the $Q_\parallel$ sheath, it ranges from approximately 2 to 4. It should be noted that the Rankine-Hugoniot conditions predict a maximum $|B|/|B_{IMF}|$ of 4 \cite{russell2016space}, which is smaller than our observed values in the inner magnetosheath. However, it should be noted that the Rankine-Hugoniot conditions assume single-fluid plasma. \citeA{fruchtman_seasonal_2023} examined the magnetic shock jump at Mars and found that many individual crossings overshot the limit of 4, suggesting that a two-fluid variation of the Rankine-Hugoniot conditions is needed at Mars. For a  \change{disucssion}{discussion} of two-fluid Rankine-Hugoniot theory, \change{please refer}{the reader is referred} to \citeA{motschmann_multiple-ion_1991} and \citeA{fahr_entropy_2015}. Our observed discrepancy is reasonable considering that magnetic fields are gradually intensified by the pile-up as they approach Mars, driven by the mass-loading effect of planetary ions \cite{Boscoboinik2023,li_ion-scale_2025}. Consequently, it suggests that the planetary ions may influence the configuration of magnetic fields in the magnetosheath.

\subsection{Solar Wind Protons}
Figure \ref{fig:hydrogen}(b1) shows the distribution of proton density, normalized to the upstream solar wind, denoted as $n_{H^+}/n_{sw}$. $n_{H^+}/n_{sw}$ peaks on the flanks, roughly within the $\pm60^o$ to $\pm75^o$ range, with a local minimum between $60^o$ and $-60^o$. $n_{H^+}/n_{sw}$ initially increases as it moves towards Mars from the BS to the middle region of the magnetosheath and then decreases as it approaches the MPB, particularly past the terminator. This suggests that the solar wind protons are initially decelerated by the BS, thus increasing their density, but subsequently deflected. The deflection could result from two factors: first, the plasma depletion layer, where strong magnetic fields near the MPB can squeeze out the plasma \cite{zwan_depletion_1976}; second, depletion caused by planetary ions.

Panels (b2)-(b3) of Figure \ref{fig:hydrogen} illustrate the asymmetry of $n_{H^+}/n_{sw}$. In the dayside region, $n_{H^+}/n_{sw}$ has a maximum value of 1.8 (1.5) in the $Q_\perp$ ($Q_\parallel$) sheath, with $n_{H^+}/n_{sw}$ in the $Q_\perp$ sheath being about 5\% higher than in the $Q_\parallel$ sheath. However, in the terminator and nightside regions, $n_{H^+}/n_{sw}$ decreases to 0.7 on both flanks, indicating weak compression in these areas. The asymmetry gradually decreases in the nightside region to just 2.8\%.

Figure \ref{fig:hydrogen}(c1) shows the magnitude of the proton velocity, normalized to the upstream solar wind speed, denoted as $|v_{H^+}|/|v_{sw}|$. Upon crossing the BS, the solar wind is significantly decelerated, especially in the subsolar region $\pm30^o$, with $|v_{H^+}|/|v_{sw}|$ a minimum of 0.4 at $+15^o$. As the solar wind propagates toward the nightside, $|v_{H^+}|/|v_{sw}|$ increases gradually (see Figure \ref{fig:hydrogen}(c2)). Overall, the deceleration of the solar wind in the magnetosheath appears to be nearly symmetric, with the asymmetry index, \textit{A}, generally below 5\% (see Figure \ref{fig:hydrogen}(c2-c3)).

 $|v_{H^+}|/|v_{sw}|$ is higher in the dayside $Q_\perp$ sheath, consistent with the behavior observed in Earth's magnetosheath \cite{dimmock_dawn-dusk_2017}, but higher in the nightside $Q_\parallel$ sheath (see Figure \ref{fig:hydrogen}c). Interestingly, this area corresponds to a kink-like magnetic structure, which naturally forms a current sheet \cite{zhang_three-dimensional_2022, romanelli_dependence_2015}. This structure is a consistent feature of the magnetosheath for IMF cone angles below $60^\circ$ and above $120^\circ$ degrees, as selected in our dataset. It can be seen in Figure \ref{fig:hydrogen}(a1) in the magnetic field vectors in the $Q_\parallel$ nightside region. This indicates that this structure likely plays a role in regulating the motion of solar wind protons.


 \begin{figure}
 \centering
 \includegraphics[width=0.99\textwidth]{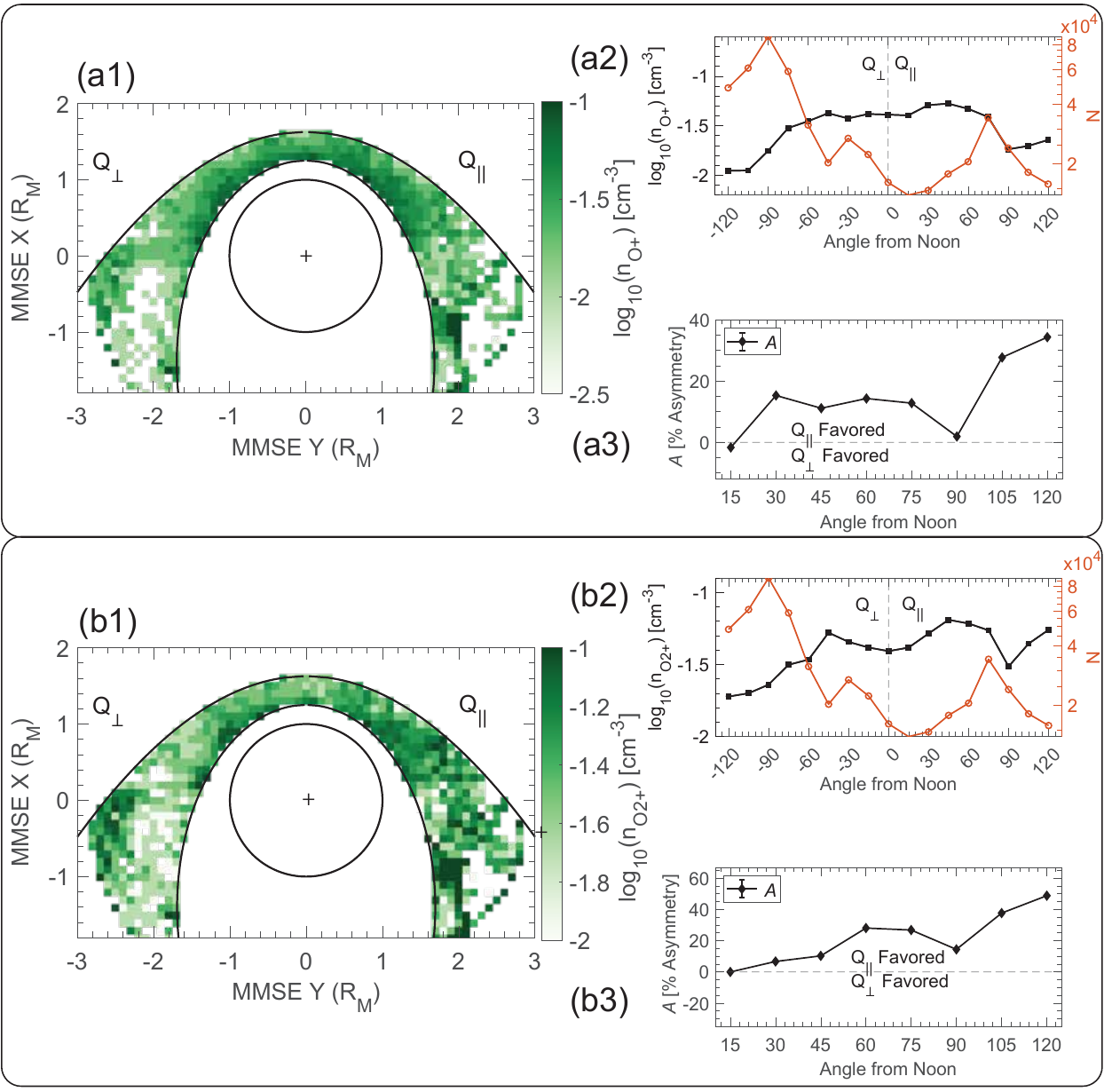}
 \caption{Following the method and format used in Figure \ref{fig:hydrogen}, panel (a) shows the number density of $O^+$, and panel (b) presents the number density of $O_2^+$.}
 \label{fig:oxygen}
 \end{figure}
 
\subsection{Planetary Ions}
Based on the results presented in Section 4.1, it is possible that planetary ions may strengthen the magnetic field and deflect solar wind protons. Therefore, it is intriguing to investigate whether the planetary ions also exhibit asymmetries within the magnetosheath. Here we focus only on planetary oxygen ions, specifically $O^+$ and $O_2^+$. Figure \ref{fig:oxygen}(a1, b1) illustrates the spatial distribution of the number densities of $O^+$ and $O_2^+$. The ion densities appear to be highly variable and are log-normally distributed. Due to this, we use the median absolute deviation, $MAD$ = median($X-\bar{X}$), where $X$ is each data point and $\bar{X}$ is the median of each bin, in place of the standard deviation when calculating the uncertainty on each bin. The number densities of $O^+$ and $O_2^+$ are generally less than 0.1 $cm^{-3}$. The typical solar wind proton density in the magnetosheath is approximately 6 $cm^{-3}$ \cite{wang_south-north_2020}. Therefore, planetary oxygen ions contribute less than 2\% to the plasma number density in the magnetosheath, but a much higher percentage of the total ion mass, as they are significantly heavier than $H^+$.

In panels (a1) and (b1) of Figure \ref{fig:oxygen}, we observe that both $O^+$ and $O_2^+$ display qualitatively similar trends. Both ions exhibit higher densities between $60^\circ$ and $-60^\circ$, which is proximal to the ionosphere and where the photo-ionization rate is the highest. Subsequently, their densities decrease toward nightside. From panels (a2), (a3), (b2), and (b3) of Figure \ref{fig:oxygen}, both $O^+$ and $O_2^+$ exhibit clear asymmetries, with higher densities on the $Q_\parallel$ flank compared to the $Q_\perp$ flank. This asymmetry increases from 0 to 20\% for $O^+$ and up to 40\% for $O_2^+$ going from the dayside to the nightside. This asymmetric trend contrasts with that of solar wind protons, which is partially caused by shock geometry. This is reasonable considering that most planetary ions do not cross the BS but originate from below the MPB. Therefore, other factors are responsible for the asymmetry. This behavior aligns with findings from kinetic hybrid simulations, which showed that the $\vec{E}\times\vec{B}$ drift causes planetary ions to migrate toward the $Q_\parallel$ flank due to conservation of momentum \cite{kallio_kinetic_2012}. As we move tailward, the concentration of planetary ions in the $Q_\parallel$ flank increases, leading to a progressively greater asymmetry. This is consistent with our results. Consequently, we suggest that the $\vec{E}\times\vec{B}$ drift is responsible for the asymmetric distribution of the planetary ions.

\section{Discussion and Conclusion}
\begin{figure}
    \centering
    \includegraphics[width=0.97\linewidth]{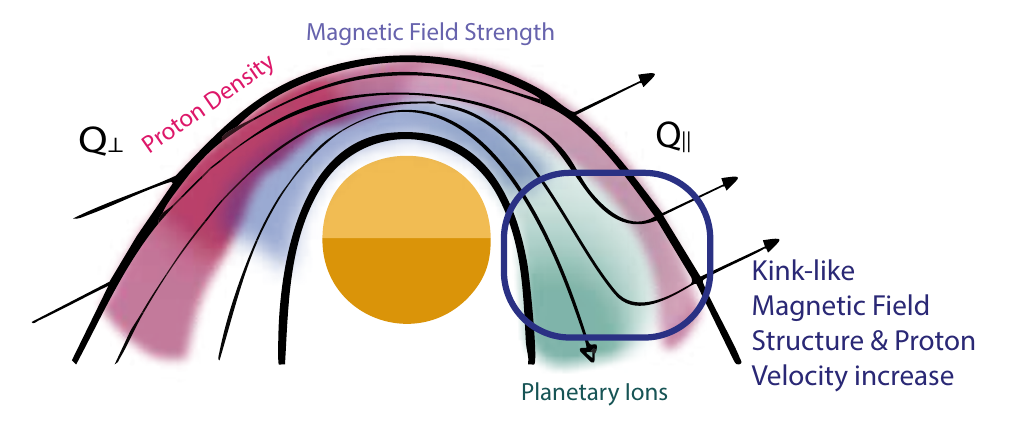}
    \caption{Summary diagram of results. The asymmetries are shaded in the region where they are strongest. The magnetic field is strongest near the MPB between $-30^\circ$ and $-60^\circ$ Proton density is highest in the $Q_\perp$ flank near $-60^\circ$. The kink-like magnetic structure is located on the $Q_\parallel$ nightside, which also corresponds to a region of increased proton speed and slight $Q_\parallel$-favored asymmetry. $O^+$ and $O_2^+$ are densest in the $Q_\parallel$ magnetosheath.}
    \label{fig:diagram}
\end{figure}

Drawing on nine years of MAVEN observations, we conducted a statistical study to investigate the asymmetries in the magnetic field and plasma between the $Q_\perp$ and $Q_\parallel$ regions of the Martian magnetosheath. We confined our study to $Z<\pm1$, $R_M$ IMF cone angles $30^\circ-60^\circ$ and $120^\circ-150^\circ$, and eliminated data that was affected by the crustal fields. Our findings indicate that, similar to Earth and Venus, Mars' magnetosheath displays pronounced asymmetries. As illustrated in Figure \ref{fig:diagram}, the primary conclusions are summarized as follows:

\begin{enumerate}
    \item The normalized magnetic field strength in the $Q_\perp$ magnetosheath region is 3\% to 14\% stronger compared to the $Q_\parallel$ magnetosheath region. This asymmetry is expected from the Rankine-Hugoniot conditions at the shock, though the compression ratio further into the sheath exceeds the theoretical maximum..
    \item In the dayside region, the solar wind proton density is 5-10\% higher in the $Q_\perp$ magnetosheath than in the $Q_\parallel$ magnetosheath. The proton speed is about 5\% higher on the $Q_\perp$ magnetosheath in the dayside region; yet, this trend reverses on the nightside, with speeds showing a 3\% increase on the $Q_\parallel$ magnetosheath. This appears consistent with the presence of the kink-like magnetic structure in the $Q_\parallel$ nightside hemisphere, which is a permanent feature of the magnetosheath for the IMF cone angles selected in our dataset.
    \item Planetary ions, $O^+$ and $O_2^+$, also exhibit significant asymmetry. Although both $O^+$ and $O_2^+$ appear nearly symmetric in the dayside region, the asymmetry increases as one moves tailward. Beyond the terminator region, the density of $O^+$ and $O_2^+$ in the $Q_\parallel$ sheath is 20-30\% higher than in the $Q_\perp$ magnetosheath. This observed asymmetry aligns with the anticipated effects of $\vec{E}\times\vec{B}$ drift.
\end{enumerate}

We found that although planetary ions do not contribute significantly to the plasma density in the magnetosheath, they likely still play an important role in its dynamics. The compression ratio of the magnetic field in the inner magnetosheath exceeds the theoretical maximum value obtained by the single-fluid Rankine-Huginiot theory, suggesting that mass loading by planetary ions likely plays a role in the amplification of the magnetic field. Additionally, the reversed asymmetry trend of solar wind proton speed past the terminator region suggests that the kink-like current sheet structure in the $Q_\parallel$ magnetosheath also plays a role in influencing plasma distribution in the magnetosheath. Within this structure, solar wind protons exhibit localized increased density, which is expected because the plasma tends to converge at the center of the current sheet.

To understand the mechanisms responsible for magnetosheath asymmetries, we compare different planetary magnetosheaths. In Earth's magnetosheath, the proton density is about 10-21\% higher in the $Q_\parallel$ magnetosheath, while the proton speed and magnetic field strength are more pronounced in the $Q_\perp$ magnetosheath, with reported asymmetries of 5-12\% and 5-23\%, respectively \cite{walsh_dawn-dusk_2012, dimmock_statistical_2013, dimmock_dawn-dusk_2016, dimmock_dawn-dusk_2017}. The studies at Earth report dawn/dusk asymmetries, but due to the nominal $45^\circ$ Parker Spiral angle \protect\cite{slavin_solar_1981}, the dawn (dusk) flank statistically corresponds to the $Q_\parallel$ ($Q_\perp$) flank. In the Venusian magnetosheath, \citeA{rojas_mata_proton_2023} found that the proton density is approximately $18\%$ higher in the $Q_\parallel$ magnetosheath while the proton speed is about $7\%$ higher in the $Q_\perp$ magnetosheath and reported no significant asymmetry in magnetic field strength. The lack of an observed $|B|$ asymmetry at Venus may be caused by poor data resolution rather than a true symmetric magnetosheath \cite{rojas_mata_proton_2023}. The asymmetries in proton speed and magnetic field strength align with those observed in Earth's and Venus's magnetosheaths. However, we observe an opposite asymmetry in the proton density in the Martian magnetosheath. Comparatively, the asymmetry in proton density in the Martian magnetosheath contrasts with observations at Earth and Venus. However, the asymmetries in proton speed and magnetic field strength align with those observed in Earth's and Venus's magnetosheaths. These results suggest that certain physical mechanisms driving these asymmetries may be universal across various planetary environments, whereas other mechanisms appear to vary depending on the specific characteristics of the surrounding space environment. For instance, the Rankine-Hugoniot conditions, commonly applied to planetary bow shocks, predict higher magnetic field strengths in the $Q_\perp$ flank, which is observed at Mars and Earth \cite{walsh_dawn-dusk_2012,dimmock_dawn-dusk_2017}. This consistent observation across Mars, Venus, and Earth supports this. Furthermore, to maintain continuity of the tangential electric field, the Rankine-Hugoniot conditions predict a shift in the flow stagnation region from the subsolar point towards the $Q_\parallel$ sheath \cite{turc_asymmetries_2020}. This results in a higher proton speed in the $Q_\perp$ magnetosheath at Earth. A similar phenomenon is observed at Mars, as illustrated in Figure \ref{fig:hydrogen}(c). Therefore, we suggest that the asymmetries in both magnetic field strength and proton speed within the planetary magnetosheath primarily stem from the shock geometry. 

The asymmetry in proton density is generally believed to be influenced by shock geometry \cite{walsh_dawn-dusk_2012}. Specifically, the BS extends further on the $Q_\perp$ side as fast magnetosonic waves propagate faster there, resulting in a thicker magnetosheath in this region. Consequently, the proton density is higher in the $Q_\parallel$ magnetosheath due to its comparatively smaller area. However, the proton density asymmetries we observed at Mars exhibit a trend opposite to that at Earth, suggesting that additional factors are involved. Compared with Earth's magnetosheath, the Martian magnetosheath exhibits two significant differences: the presence of planetary ions and more pronounced kinetic effects due to its small extent. Given that planetary oxygen ions exhibit an opposite trend to protons, it is plausible that the proton asymmetry at Mars is related to these ions. However, the factors that cause the asymmetric proton density distribution remain unclear and are beyond the scope of this paper, requiring further investigation.

\section*{Data Availability Statement}
The research described in this manuscript utilizes publicly available data from the MAVEN mission, including
data from the SWIA, MAG, and STATIC instruments \cite{connerney2023maven,10.17189/1414182,10.17189/1517741}. The data analysis was performed using the irfu-matlab software package \cite{khotyaintsev_irfu-matlab_2024}.

\acknowledgments
We acknowledge helpful discussions with Gwen Hanley. This work was partially supported by the MAVEN Project, NASA grants 80NSSC23K1125, 80NSSC23K0911, 80NSSC24K1843, and the Massachusetts Space Grant Consortium (MASGC) awards to Boston University.

%

%


%
%
%
%
%

\end{document}